\documentclass[10pt,pra,twocolumn,superscriptaddress,showpacs,floatfix]{revtex4-2}
\usepackage{graphics}
\usepackage{graphicx}
\usepackage{bm}
\usepackage[dvipsnames]{xcolor}

\begin{document}
\title[Modeling acceleration without photon pair creation]{Modeling acceleration without photon pair creation}

\author{Sara Kanzi}
\address{Faculty of Engineering, Final International University, North Cyprus Via Mersin 10, Kyrenia, 99370, Turkey}
\author{Daniel Hodgson}
\address{School of Physics and Astronomy, University of Leeds, Leeds, UK, LS2 9JT}
\address{School of Mathematical and Physical Sciences, University of Sheffield, Sheffield S3 7RH, United Kingdom}
\author{Almut Beige}
\address{School of Physics and Astronomy, University of Leeds, Leeds, UK, LS2 9JT}
\vspace{10pt}

\begin{abstract}
The Unruh effect predicts that a uniformly accelerating observer will find themselves in a thermal bath of photons, where a resting observer would not see any. However, since Maxwell's equations hold everywhere, the observers cannot tell who is resting and who is accelerating by looking only at the electromagnetic field. The only difference between different reference frames is that they use different spacetime coordinates; otherwise the formalism for the description of light must be the same for all observers. To provide new insight into the validity of the Unruh effect, this paper generalises a recent physically-motivated quantisation of the electromagnetic field to different reference frames. Afterwards, we demonstrate that, in position space, an accelerating observer simply experiences a constantly changing Doppler effect, while all reference frames share a common vacuum.
\end{abstract}

%
%
%
\maketitle
%

\section{Introduction}

It is a widely held belief that the modeling of the quantised electromagnetic (EM) field seen by an accelerating observer cannot be done without involving Bogoliubov transformations \cite{Cri}. This transformation replaces the observer's photon annihilation operators in the resting reference frame with a mixture of photon annihilation and creation operators. Consequently, the resting observer and the accelerating observer do not share a common vacuum. Instead, the accelerating observer is expected to encounter photon pairs, even when the resting observer does not see any particles. The resulting Unruh effect, which is also known as the Fulling-Davies-Unruh effect \cite{Unruh,Unruh2,Unruh3}, seems to be a mathematical necessity \cite{Rosabal:2019nrf,Ramakrishna:2023ntp,Berra-Montiel:2016xnn,Volovik:2022cqk,Rosabal:2018hkx,Kialka:2017ubk,Buchholz:2012wr, Buchholz:2014jta,Maybee}. To provide at least some physical reasoning, the Unruh effect has been studied over the years from a wide range of perspectives. For example, some authors invoke arguments from thermodynamics or highlight the presence of boundary conditions (cf.~e.g.~Refs.~\cite{Birrell,Ref1,Ref2,Muk}); others focus on detector models \cite{Unruh2,UDD,UDD2,UDD3}. However, intuitively there there does not seem to be anything quantum about moving through vacuum. 

\begin{figure}[t]
\centering\includegraphics[width=0.45 \textwidth]{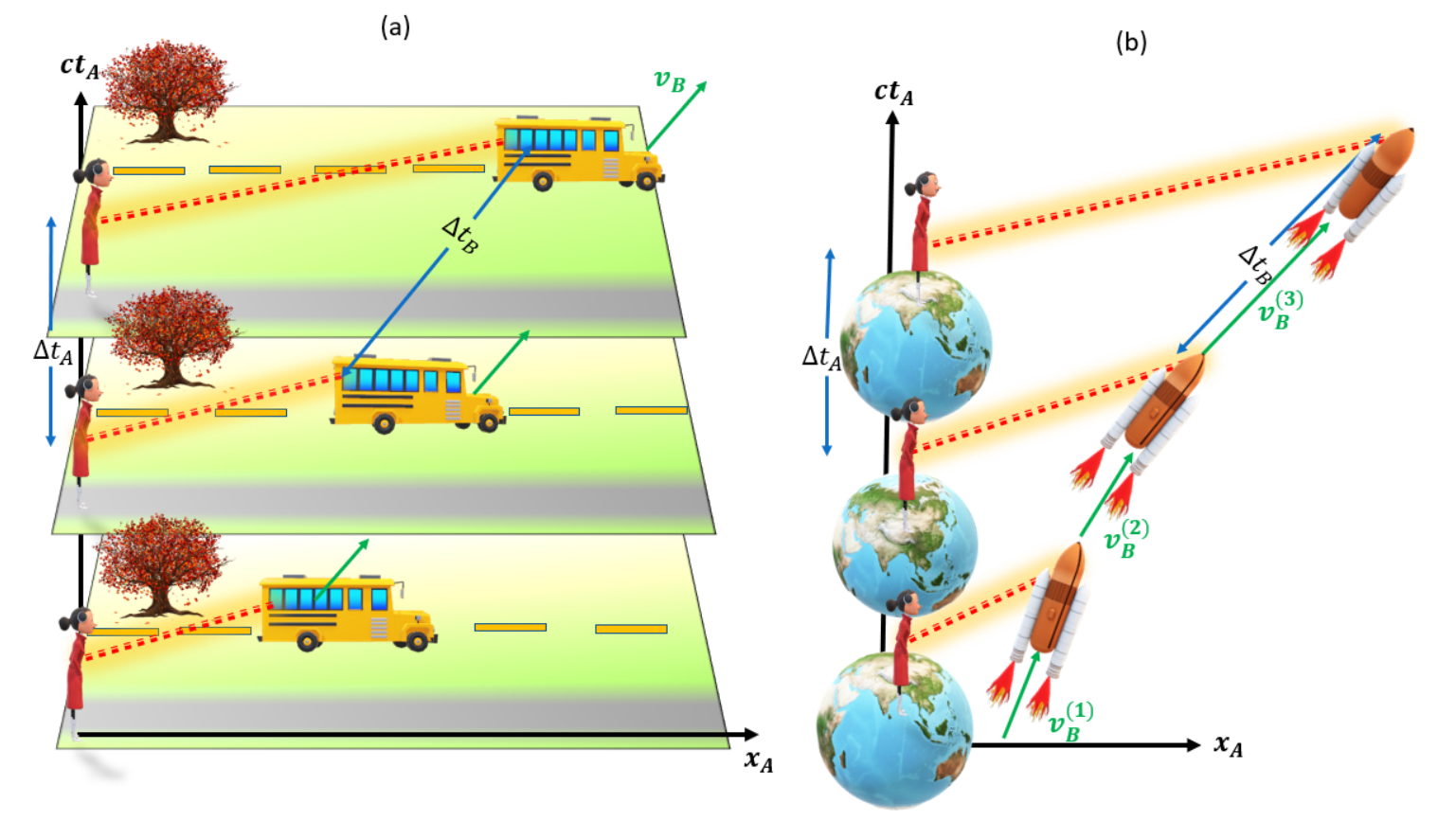}
\caption{(a) A spacetime diagram illustrating the Doppler effect. Here, a moving observer (Bob) travels at a constant velocity $v_{\rm B}$ (green line) away from a resting observer (Alice). In addition, we assume that Alice sends light signals to Bob at intervals of $\Delta t_{\rm A}$, which he receives at constant time intervals $\Delta t_{\rm B}$. (b) A spacetime diagram illustrating the Unruh effect. Now Bob moves with a time-dependent velocity $v_{\rm B} (t_{\rm A})< c$ (green line) away from Alice. Notice that Bob's dynamics look the same in both figures during a short time interval $\Delta t_{\rm B}$ as his trajectory can be approximated by a collection of short straight lines at velocity $v_{\rm B}^{(m)}$ with $m=1,2,3,...$\,.}\label{FIGUREs1}
\end{figure}  
 
To confirm the Unruh effect, experiments have studied the interaction between the free radiation field and accelerating charged particles (which play the role of the accelerating observer), but remained inconclusive regarding the possible creation of photon pairs from the vacuum \cite{CERN,Lynch:2019hmk,Gregori:2023tun,Lynch:2023zll}. In addition, analogue gravity experiments have been performed \cite{Nicolaevici:2014ela, Leonhardt}. These, however, are not independent confirmations of photon pair creation in the reference frame of an accelerating observer, since they are designed to exhibit a certain behavior. The Unruh effect is therefore still the subject of an ongoing debate. The question that remains is whether the origin of the problem are paradoxical physical processes or issues related to the original formulation of quantum field theory. In this paper, we show that it is possible to model acceleration without photon pair creation when describing the quantised EM field using a recently-introduced local photon approach \cite{Jake,Hodgson2022,Gabe}. Eventually, the question of which description of the EM field is the most accurate needs to be answered by experiments.

An effect that is closely related to the Unruh effect, but much better understood, is the relativistic Doppler effect which concerns two observers moving at constant relative speed \cite{Michel2021,Huang2010}.
When observing the same light signal, the signal's spectrum appears differently in the reference frames of the resting and of the moving observer. To see that this is indeed the case, consider a stationary observer called Alice who sends narrow light pulses at regular time intervals to a moving observer called Bob \cite{Hodgson:2024mfs}, as illustrated in Fig.~\ref{FIGUREs1}(a). From special relativity, we know that the speed of light $c$ is the same everywhere. Hence, the spacetime coordinates $(x_{\rm A},t_{\rm A})$ and $(x_{\rm B},t_{\rm B})$ used by Alice and Bob to describe the same narrow light pulse are in general not the same. Taking the transformation between these spacetime coordinates into account that, we find that the corresponding local photon annihilation operators $a_{s \lambda}(x_{\rm A},t_{\rm A})$ and $b_{s \lambda}(x_{\rm B},t_{\rm B})$ in the Heisenberg picture transform such that \cite{Hodgson:2024mfs}
\begin{eqnarray} \label{A1}
a_{s \lambda}(x_{\rm A},t_{\rm A}) &=& \zeta_{\rm AB} \, b_{s \lambda}(x_{\rm B},t_{\rm B})
\end{eqnarray}
when referring to the same narrow light pulse. Here $\zeta_{\rm AB}$ is a constant that depends only on Bob's velocity with respect to Alice. Eq.~(\ref{A1}) guarantees that both operators $a_{s \lambda}(x_{\rm A},t_{\rm A})$ and $b_{s \lambda}(x_{\rm B},t_{\rm B})$ obey bosonic commutator relations. When transforming the local operators in Eq.~(\ref{A1}) into momentum space, $\zeta_{\rm AB}$ yields the expected relativistic Doppler shift \cite{Hodgson:2024mfs}.

Taking this into account, the particle generation predicted by the Unruh effect seems even more mysterious, since the trajectory of the accelerating observer can be approximated by a collection of short time intervals during which the observer moves, as in the Doppler scenario, at constant speed (cf.~Fig.~\ref{FIGUREs1}(b)). Moreover, one would expect that Eq.~(\ref{A1}) holds also for an accelerating observer, since narrow light pulses should appear narrow in all reference frames. As we shall see in the following, the only difference between a moving observer and an accelerating observer is that, in the latter case, the constant $\zeta_{\rm AB}$ depends not only on the current velocity of Bob, but on his whole trajectory within Alice's reference frame. Most importantly, notice that Eq.~(\ref{A1}) is not a Bogoliubov transformation but preserves the vacuum in all reference frames. The results presented in this paper therefore suggest that the Unruh effect \cite{Unruh, Unruh2,Unruh3} could be an oversight.

The traditional approach to obtaining a quantum description of a non-interacting classical theory is canonical quantization \cite{Birrell}. Although this approach is mathematically elegant, the physical meaning of its state vectors and operators is not always clear. For example, in the case of quantum field theory, the vector potential of the quantised EM field is gauge-dependent. Nevertheless, it is used to define the annihilation and creation operators of actual photons. To overcome the dependence of the quantum formalism on a specific gauge, Bennett {\em et al.}~\cite{Bennett} proposed to quantise light taking a more physically motivated approach and to promote the distinguishable classical solutions of Maxwell's equations in free space to pairwise orthogonal quantum states. When considering narrow light pulses with different polarisations and different directions of propagation, which travel at the speed of light and therefore solve Maxwell's equations, we obtain the local photon annihilation operators $a_{s \lambda}(x_{\rm A},t_{\rm A})$ in Eq.~(\ref{A1}) \cite{Jake,Hodgson2022,Gabe}.

While the formalisms arising from canonical and from physically-motivated quantisations of the EM field are formally the same, the local photon approach of Refs.~\cite{Jake,Hodgson2022,Gabe} reveals a larger Hilbert space than the state space that is usually considered in quantum field theory. In this paper, we study the relevance of this observation for the predictions of the Unruh effect \cite{Unruh,Unruh2,Unruh3}. As in our previous treatment of the relativistic Doppler effect 
\cite{Hodgson:2024mfs}, we consider local rather than monochromatic photons as the basic building blocks of the quantised EM field. This allows us to directly exploit several physical principles. For example, the {\em light principle} tells us that light propagates at constant speed, $c$, independently of the motion of its source and its receiver \cite{Einstein}. The only difference between Alice and Bob is that they use different spacetime coordinates to describe events. However, the {\em principle of relativity} tells us that, nevertheless, they experience the same physical laws.

To simplify the derivation of the constant $\zeta_{\rm AB}$ in Eq.~(\ref{A1}) for the case of an accelerating observer, we notice that not only the speed $c$ but also the direction of propagation $s$ of light is invariant under reference frame transformations. This allows us to replace the spacetime coordinates $(x_i,t_i)$ that define the trajectory of narrow light pulses in the reference frames of Alice and Bob ($i= {\rm A}, {\rm B}$) by a single natural coordinate $\chi_i \equiv x_i-sct_i$ \cite{Hodgson:2024mfs}, if $s = \pm 1$ refers to light propagating to the right and to the left, respectively. The coordinates $\chi_i$ are constant for all points belonging to the light line of a narrow light pulse. 
Using this notation, Eq.~(\ref{A1}) simplifies to
\begin{eqnarray} \label{A2}
a_{s \lambda}(\chi_{\rm A}) &=& \zeta_{\rm AB} \, b_{s \lambda}(\chi_{\rm B})
\end{eqnarray}
with $a_{s \lambda}(\chi_{\rm A})$ and $b_{s \lambda}(\chi_{\rm B})$ representing bosonic operators. As we can see from this relation, all we need to know to determine $\zeta_{\rm AB}$ is how the natural coordinates $\chi_{\rm A}$ and $\chi_{\rm B}$ transform when referring to the same light line. To determine this relationship, we assume again that Alice sends narrow light pulses to Bob at regular time intervals. For simplicity, we place both observers at the origins of their respective spacetime diagrams and assume that they remain there at all times. In addition, we assume that Bob moves with a  time-dependent velocity $v_{\rm B}(t_{\rm A})$ with $ |v_{\rm B}(t_{\rm A})| < c$ along a trajectory in Alice's spacetime diagram. As both Alice and Bob progress along time-like trajectories in their respective Cartesian coordinate systems, they intersect the trajectory of each narrow light pulse only once. Hence, there is a single-valued relationship between the natural coordinates $\chi_{\rm A}$ and $\chi_{\rm B}$ corresponding to the same light-like trajectory but viewed from different reference frames.

In the remainder of this paper, we first review our local photon approach to the quantisation of light \cite{Jake,Hodgson2022,Gabe,Sarben}. In addition, we establish the relationship between Alice's and Bob's natural coordinates $\chi_{\rm A}$ and $\chi_{\rm B}$ for a given trajectory of the accelerating observer (Bob) with respect to the Cartesian coordinates of the stationary observer (Alice). Afterwards, we use our results to identify the transformation of the local photon annihilation and creation operators from one reference frame to the other. The main purpose of this paper is to show that it is indeed possible to model the quantised EM field of a non-uniformly moving observer in a way that is consistent with both quantum physics and with special relativity without having to replace the vacuum in one reference frame with a thermal bath in the other.

\section{Methods}

\subsection{An alternative quantisation of the EM field} \label{oma4}

This paper aims to study the Unruh effect using a recently introduced  local photon approach \cite{Jake,Hodgson2022,Gabe}. In this section, we review the state space, the Hamiltonian, and the basic observables of this approach to quantum electrodynamics, which is based on a physically-motivated quantisation of the EM field \cite{Bennett} in position space. As we shall see below, the resulting quantum formalism can easily accommodate the basic postulates of special relativity. In contrast to standard formulations of quantum field theory, our Hamiltonian is locally acting, unbounded, and does not allow initially localised wave packets to disperse at superluminal speeds, as is the case for bounded Hamiltonians \cite{Hegerfeldt,Mal,AMC}.

\subsubsection{Local Photons}

To describe wave packets of light propagating along the $x$ axis, we promote the classically distinguishable solutions of Maxwell's equations in position space to pairwise orthogonal single-photon quantum states and also allow for the presence of many photons. More concretely, we decompose photonic wavepackets in the following into localised field excitations, so-called {\em bosons localised in position} (blips). In the Heisenberg picture, we denote their annihilation operators by $a_{s\lambda}(x,t)$ with $x \in (-\infty,\infty)$ characterising their positions and with $s = \pm 1$ and $\lambda = {\sf H},{\sf V}$ representing their direction of propagation and their polarisation, respectively. As we know from special relativity, wave packets of any shape solve Maxwell's equations in 1+1 dimensions as long as they travel at constant speed along light-like (null) trajectories in the spacetime diagram. Hence, in the absence of interactions, the expectation values of all annihilation operators $a_{s\lambda}(x,t)$ where $x$ and $t$ give the same value of the natural coordinate $\chi = x -sct$ have the same expectation values as a consequence of their dynamics \cite{Jake,Hodgson2022,Gabe}.
Hence
\begin{eqnarray}
a_{s\lambda}(x,t) &=& a_{s\lambda}(x-sct,0) \, .
\end{eqnarray}
For simplicity, we therefore replace all $a_{s\lambda}(x,t)$ operators associated with the same $\chi$ in the following by a single annihilation operator $a_{s\lambda}(\chi)$ and refer to $\chi$ as the natural coordinate.

Linear optics experiments clearly demonstrate that photons are bosonic particles \cite{Kok,Haw1,Hawton2021}, and hence
\begin{eqnarray}
\label{comm1}
\big[a_{s\lambda}(\chi),a^{\dagger}_{s'\lambda'}(\chi')\big]
&=& \delta_{ss\prime}\,\delta_{\lambda\lambda^{\prime}}\,\delta(\chi-\chi') \, , 
\end{eqnarray}
while all other operators commute. Proceeding as in Refs.~\cite{Jake,Hodgson2022,Gabe} and using the above commutation relations, it can be demonstrated that the {\em dynamical} Hamiltonian 
\begin{eqnarray}
\label{Hdyn}
H_{\rm dyn} &=& \sum_{s=\pm 1}\sum_{\lambda={\sf H},{\sf V}}\int_{-\infty}^{\infty}\text{d}\chi \, \hbar sc\, a^\dagger_{s\lambda}(\chi) \frac{\partial}{\partial \chi}a_{s\lambda}(\chi) ~
\end{eqnarray}
propagates wave packets at the speed of light in their specified direction $s$. The above Hermitian Hamiltonian is not positive. Due to the presence of the parameter $s$, it has an equal number of positive and negative eigenvalues.

\subsubsection{Monochromatic photons}

The dynamical Hamiltonian $H_{\rm dyn}$ can be diagonalised by introducing monochromatic excitations with annihilation operators 
\begin{eqnarray} \label{mom1}
\widetilde{a}_{s\lambda}(k) &=& \int_{-\infty}^{\infty}\frac{\text{d}\chi}{\sqrt{2\pi}} \, {\rm e}^{-{\rm i}sk\chi} \, a_{s\lambda}(\chi)\, ,
\end{eqnarray}
which are superpositions of all blip annihilation operators $a_{s\lambda}(\chi)$. To obtain a complete set of operators, we need to consider $k \in (-\infty, \infty)$, while $s = \pm 1$ and $\lambda = {\sf H},{\sf V}$ as in the previous subsection. Since the $\widetilde{a}_{s\lambda}(k)$ and the $a_{s\lambda}(\chi)$ relate to each other via a Fourier transform (cf.~Eq.~(\ref{mom1})), they are both bosonic and 
\begin{eqnarray}
\label{comm1k}
\big[\widetilde{a}_{s\lambda}(k),\widetilde{a}^{\dagger}_{s'\lambda'}(k')\big]
&=& \delta_{ss\prime}\,\delta_{\lambda\lambda^{\prime}}\,\delta(k-k') \, . 
\end{eqnarray}
The above description of photons is different from the standard description, where positive and negative $k$ describe right and left moving photons, respectively, since it involves a doubling of the usual Hilbert space of the quantised EM field. The additional parameter $s$ allows us to circumvent previous no-go theorems \cite{NW,Bia,Sipe} regarding the impossibility of representing photonic wave packets in position space \cite{Cook,Cook2,Mandel,Jordan,Pendry,Hawton,Fan,Babaei2017,Casimir2}.

When substituting Eq.~(\ref{mom1}) into Eq.~(\ref{Hdyn}), the dynamical Hamiltonian simplifies to its diagonal form
\begin{eqnarray}
\label{Hdyn2}
H_{\rm dyn} &=& \sum_{s=\pm 1}\sum_{\lambda={\sf H},{\sf V}}\int_{-\infty}^{\infty}\text{d}k\; \hbar c k \,\widetilde{a}^\dagger_{s\lambda}(k)\widetilde{a}_{s\lambda}(k) \, .
\end{eqnarray}
This Hamiltonian bears a strong resemblance to the standard Hamiltonian of the quantised EM field, which also represents its energy observable. However, its eigenvalues extend into the negative region and the operator is no longer bound from below. As Maxwell's equations are time-reversible, eigenstates of the Schr\"odinger equation with both positive and negative eigenvalues must be included, and therefore both positive and negative frequencies $\omega = ck$ are required to ensure that all possible single-photon solutions of Maxwell's equations in position space are considered (cf.~Eq.~(\ref{comm1k})). 

As energy is always strictly positive, the dynamical Hamiltonian and the energy observable cannot be the same. Nevertheless, since the energy of the EM field is conserved in the absence of interactions, the dynamical Hamiltonian and energy observables must have a mutual set of eigenstates such that they commute. In the following, we therefore define the energy observable $H_{\rm eng}$ as in Ref.~\cite{Gabe} such that
\begin{eqnarray}
\label{energy1}
H_{\rm eng} &=& \sum_{s=\pm 1}\sum_{\lambda={\sf H},{\sf V}}\int_{-\infty}^{\infty}\text{d}k\; \hbar c |k| \,\widetilde{a}^\dagger_{s\lambda}(k)\widetilde{a}_{s\lambda}(k)\, .
\end{eqnarray}
This equation shows that each photon with wave number $k$ contributes the energy $\hbar c|k|$ to the total energy of the EM field. This observation is consistent with energy conservation given that an atom with transition frequency $\omega > 0$ loses the energy $\hbar \omega$ upon the emission of a photon. Notice that the negative-frequency photons in the above equations simply emerge here, since they occur naturally in the spectrum of localised objects, as defined by the Fourier transform in Eq.~(\ref{mom1}).

\subsubsection{EM field observables}

The local excitations of the quantised EM field, or blips, studied in the previous subsections are pairwise orthogonal and have unique and well-defined positions. Moreover, as shown in Refs.~\cite{Jake,Hodgson2022}, they are carriers of non-local complex electric field amplitudes. The observables ${\mathcal E}_{s \lambda}(\chi)$ of these amplitudes are non-local linear superpositions of the blip annihilation operators $a_{s\lambda}(\chi)$ and \cite{Gabe}
\begin{eqnarray}
\label{Fields1}
{\mathcal E}_{s \lambda}(\chi)
&=& \int_{-\infty}^{\infty} {\rm d}\chi' \, R(\chi,\chi^{\prime})\,
a_{s \lambda}(\chi^{\prime})
\end{eqnarray}
with the non-local distribution $R(\chi,\chi')$ given by
\begin{eqnarray}
\label{d1}
R(\chi,\chi^{\prime}) &=& -\sqrt{\frac{\hbar c}{4\pi\varepsilon A}} \cdot \frac{1}{|\chi-\chi^\prime|^{3/2}} \, .
\end{eqnarray}
Here $\varepsilon$ and $A$ denote the vacuum permittivity and the area occupied by the electric fields in the $y$-$z$ plane. The observable for the real part of the electric field amplitudes is simply given by $({\mathcal E}_{s \lambda}(\chi) + {\rm H.c.})/2$. Like their local carriers,
the field observables travel at the speed of light along the $x$ axis and therefore satisfy Maxwell's equations. Most importantly, Eq.~(\ref{Fields1}) shows that the electric  field amplitude ${\mathcal E}_{s \lambda}(\chi)$ is non-local and can be influenced by blips at positions $\chi \neq \chi'$.

When replacing the blip annihilation operators $a_{s \lambda}(\chi^{\prime})$ in Eq.~(\ref{Fields1}) by the $\widetilde{a}_{s \lambda}(k)$'s and taking into account that  
\begin{eqnarray}
\int_{-\infty}^{\infty} {\rm d}\chi' \, 
{\rm e}^{-{\rm i}sk (\chi-\chi')} \, 
R(\chi,\chi^{\prime})
&=& \sqrt{\frac{2\hbar c |k|}{\varepsilon A}}  \, ,
\end{eqnarray}
the observables of the complex electric field amplitudes assume a more familiar expression. In agreement with Ref.~\cite{Gabe}, we find that these observables can also be written as
\begin{eqnarray}
\label{Fields1extra}
{\mathcal E}_{s \lambda}(\chi)
&=& 
\int_{-\infty}^{\infty} {\rm d}k \, 
\sqrt{\frac{\hbar c |k|}{\pi\varepsilon A}}  \, {\rm e}^{{\rm i}sk \chi} \, 
\widetilde{a}_{s \lambda}(k) \, .
\end{eqnarray}
The distribution $R(\chi,\chi^{\prime})$ in Eq.~(\ref{d1}) ensures that the energy observable $H_{\rm eng}$ of the EM field in Eq.~(\ref{energy1}) can also be written as \cite{Gabe}
\begin{eqnarray}
\label{S34}
H_{\rm eng} &=&  \int_{-\infty}^{\infty} {\rm d} \chi \, h_{\rm eng}(\chi) 
\end{eqnarray}
with the energy density $h_{\rm eng}(\chi)$ given by 
\begin{eqnarray}
\label{S34b}
\hspace*{-0.5cm}
h_{\rm eng}(\chi) &=& \frac{A\varepsilon}{2} 
\sum_{s=\pm 1} \sum_{\lambda = {\sf H}, {\sf V}} {\mathcal E}_{s \lambda}^\dagger(\chi) \cdot {\mathcal E}_{s \lambda}(\chi) \, . 
\end{eqnarray}
Although the above equations are not needed here to obtain new insights into the Unruh effect, we have added them to highlight that our local photon description of the quantised EM field is analogous to its standard description in quantum electrodynamics apart from the above described doubling of the Hilbert space.

\subsection{Coordinate transformations between inertial reference frames} 

\begin{figure}[t]
\centering\includegraphics[width=0.45 \textwidth]{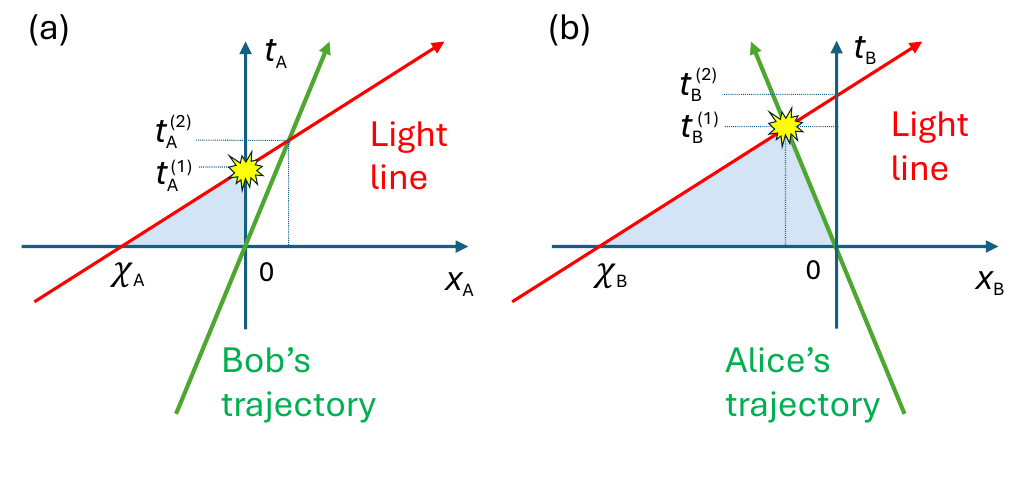}
\caption{(a) Alice's spacetime diagram, with coordinate axes $x_{\rm A}$ and $t_{\rm A}$, showing Bob's time-like trajectory (green) for the case where Bob moves at constant speed $v_{\rm B}> 0$ away from her. Alice and Bob both met at the origin of their respective coordinate systems at an initial time $t_{\rm A} = t_{\rm B} = 0$. The red line marks the light-like trajectory (red) of a narrow light pulse caused by an explosion at Alice's position at a time $t_{\rm A}^{(1)}$ and arriving at Bob's location at $t_{\rm A}^{(2)}$.
		(b) Illustration of the same scenario in Bob spacetime diagram with coordinate axes $x_{\rm B}$ and $t_{\rm B}$ showing the trajectory of the light pulse (red) caused by an explosion at time $t_{\rm B}^{(1)}$ and reaching him at $t_{\rm B}^{(2)}$. Alice's trajectory is marked in green. Both figures can be used to derive Eq.~(\ref{sk0}) and subsequently the transformation between the natural coordinates $\chi_{\rm A}$ and $\chi_{\rm B}$ that correspond to the same event in Eq.~(\ref{S01}).} \label{FIGUREb1}
\end{figure}

The above quantisation of the EM field in position space can be carried out in any reference frame. To find out how the  blip annihilation operators $a_{s \lambda}(\chi_{\rm A})$ and $b_{s \lambda}(\chi_{\rm B})$ of two observers, Alice and Bob, relate to each other, we take into account that blips remain blips in all reference frames \cite{Hodgson:2024mfs}. However, the natural coordinates $\chi_{\rm A}$ and $\chi_{\rm B}$ which characterise the same light line are generally not the same, even when referring to the trajectory of the same narrow light pulse, since Alice and Bob experience space and time differently. Events that occur simultaneously in Alice's reference frame, but at different positions, occur at different times in Bob's reference frame. Analogously, events that occur at the same location at different times in one reference frame are spatially separated in the other. As illustrated in Fig.~\ref{FIGUREb1}, the direction of propagation of light is invariant under reference frame transformations.

For simplicity, we first consider the scenario where Bob moves at a constant speed $v_{\rm B}>0$ away from Alice. Without restrictions, we assume in the following that the origins of Alice's and Bob's spacetime diagrams coincide such that Alice's light-like trajectory $\chi_{\rm A} = 0$ corresponds to the $\chi_{\rm B} = 0$ trajectory in Bob's reference frame. In order to establish the relationship between $\chi_{\rm A}$ and $\chi_{\rm B}$, we assume that Alice triggers an explosion at her position at a time $t_{{\rm A}}^{(1)}$ in her own reference frame which corresponds to the time $t_{{\rm B}}^{(1)}$ in Bob's reference frame. The explosion creates a narrow light pulse which reaches Bob at a time $t_{{\rm B}}^{(2)}$ which corresponds to the time $t_{{\rm A}}^{(2)}$ in Alice's reference frame. We can then see from Fig.~\ref{FIGUREb1} that the natural coordinates $\chi_{\rm A}$ and $\chi_{\rm B}$ that characterise the light line created by the explosion satisfy the relations
\begin{eqnarray} \label{sk0}
&& \chi_{\rm A} = - c t_{\rm A}^{(1)} = - (c - v_{\rm B}) t_{\rm A}^{(2)} \, , \nonumber\\
&& \chi_{\rm B} = - (c + v_{\rm B}) t_{{\rm B}}^{(1)} = - c t_{\rm B}^{(2)} \, . ~~
\end{eqnarray}
The above equations apply since $\chi_{\rm i}$ coincides with $-ct_{i}$ where $t_{i}$ is the time at which the light pulse arrives at $x_{\rm i}=0$. Moreover, notice that we obtain the same equations if we assume that Alice moves at a velocity $-v_{\rm B}$ within Bob's reference frame, as one would expect from the principle of relativity.

Next, we notice that the principle of relativity also implies that a possible time dilation factor $\gamma$ is the same in both reference frames \cite{Hodgson:2024mfs}. Hence
\begin{eqnarray} \label{dilation}
 {t_{{\rm B}}^{(1)} \over t_{{\rm A}}^{(1)}} = {t_{{\rm A}}^{(2)} \over t_{{\rm B}}^{(2)}} &=& \gamma \, .
\end{eqnarray} 
To calculate the proportionality factor $\gamma$, we combine Eqs.~(\ref{sk0}) and (\ref{dilation}). Proceeding in this way, we can show that
\begin{eqnarray}
\label{S00}
\chi_{\rm B} &=& \gamma \left( 1+ \beta \right) \chi_{\rm A} 
\end{eqnarray}
where $\beta=v_{\rm B}/c$ and $\gamma = (1 - \beta^2)^{-1/2}$. This relation only applies when $v_{\rm B}>0$ and $s =1$ or, using symmetry, when $v_{\rm B}<0$ and $s=-1$. To determine the equivalent relation when $v_{\rm B}>0$ and $s = -1$, Bob must send a light pulse to Alice, as any left-travelling light pulse emitted by Alice will not meet Bob. In this instance, by using the principle of relativity, which means that we cannot determine whether Alice is moving away from Bob or Bob is moving away from Alice, it must be that $\chi_{\rm A}$ and $\chi_{\rm B}$ are related by the transformation
\begin{eqnarray}
\label{S02}
\chi_{\rm A} &=& \gamma \left( 1+ \beta \right) \chi_{\rm B}\, , 
\end{eqnarray}
which is analogous to Eq.~(\ref{S00}) with the roles of the observers reversed. Results (\ref{S00}) and (\ref{S02}) can then be combined to show that
\begin{eqnarray} \label{S01}
\chi_{\rm B} &=& \gamma \left( 1+ s\beta \right) \chi_{\rm A} 
\end{eqnarray}
which applies for all values of $v_{\rm B}$ and $s$. The above relations apply for all values of $\chi_{\rm A}$ and $\chi_{\rm B}$ as all light-like worldlines must cross the trajectories of both observers exactly once.
   
Now suppose that Alice creates little explosions that send out narrow light pulses from her location at $x_{\rm A} = 0$ at regular time intervals $\Delta t_{\rm A}$. These light pulses are received at regular time intervals  $\Delta t_{\rm B}$ in Bob's reference frame. In this case, the corresponding changes in natural coordinates $\Delta \chi_{\rm A}$ and $\Delta \chi_{\rm B}$ between two subsequent light pulses in Alice's and Bob's reference frames, respectively, are not equal due to the relative motion of the observers. However, using spatial and time-translational symmetries, the ratio $\Delta \chi_{\rm B}/\Delta \chi_{\rm A}$ must be the same as the ratio $\chi_{\rm B}/\chi_{\rm A}$ of the natural coordinates $\chi_{\rm B}$ and $\chi_{\rm A}$ in Eq.~(\ref{S01}). This ratio depends only on Bob's velocity $v_{\rm B}$ in Alice's reference frame and on the direction of propagation $s$ of the narrow light pulses.

In special relativity, the spacetime coordinate transformations between two inertial reference frames are described by the Lorentz transformations, which are considered to be fundamental. In the above, Eq.~(\ref{S01}) expresses the Lorentz transformations in 1+1 dimensions in terms of the natural coordinates $\chi_{\rm A}$ and $\chi_{\rm B}$ of a stationary and a moving observer with a constant relative velocity. It is therefore not surprising that the variable transformation in Eq.~(\ref{S01}) can be used to derive the relativistic Doppler effect \cite{Hodgson:2024mfs}, which occurs in exactly the same scenario.

\subsection{Coordinate transformations between an inertial and non-inertial reference frames} \label{Sec:Generaltransformations} 

\begin{figure}[t]
\centering\includegraphics[width=0.45 \textwidth]{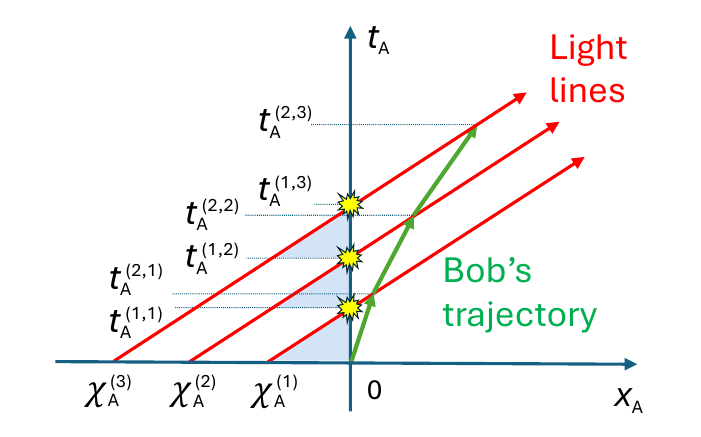}
\caption{A diagram illustrating the discretisation of Bob's trajectory (green) into short periods of inertial motion. A rapid succession of explosions by Alice sends a train of narrow light pulses to Bob at regular time intervals from her position at $x_{\rm A} = 0$ (vertical axis). Bob receives one pulse at the end of each period. As Bob is accelerating, the light beams due to the explosions reach him at different positions and times. The first number in the time superscript indicates the trigger time by Alice and the arrival time at Bob, and the second number indicates the explosion number.} \label{FIGURE001a}
\end{figure} 

In general, the relative motion between the two observers Alice and Bob can be non-uniform \cite{Ostapchuk:2011ud,Soares:2020qlw,Ding:2021nrg}. In the following, we again treat Alice as the stationary observer; however, this time we allow Bob to accelerate away from her in a possibly time-dependent way. To determine the relationship between the natural coordinates $\chi_{\rm A}$ and $\chi_{\rm B}$ in this case, we assume again that Alice triggers multiple explosions and sends narrow light pulses at regular time intervals $\Delta t_{\rm A}$ to Bob. In addition, we introduce a coarse graining and, as illustrated in Fig.~\ref{FIGURE001a}, segment Bob's continuous trajectory in Alice's spacetime diagram into a collection of short linear intervals where he moves with a constant velocity $v_{\rm B}(\chi_{\rm A})$. To take advantage of the results in the previous subsection, we assume that Bob's velocity changes whenever he crosses a light line belonging to one of the narrow light pulses sent by Alice. However, notice that in the limit of very small $\Delta t_{\rm A}$, Bob's trajectory through Alice's spacetime diagram matches his actual trajectory. 

\begin{figure}[t]
\centering\includegraphics[width=0.45 \textwidth]{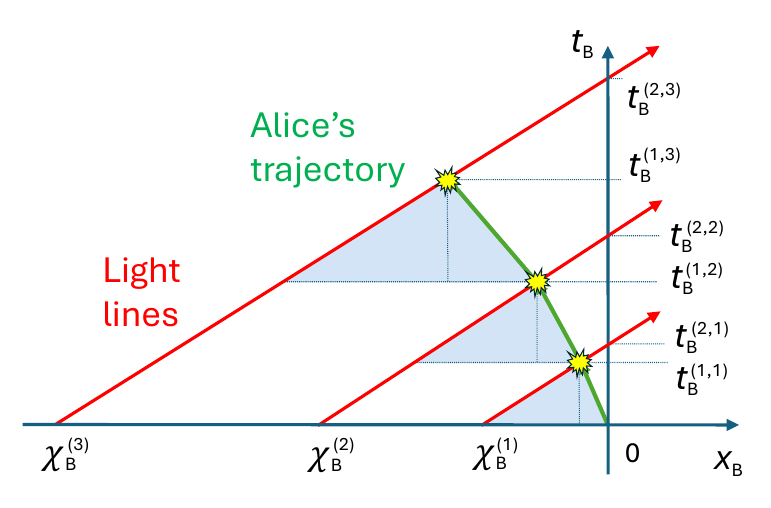}
\caption{An illustration of the explosions triggered by Alice from Bob's perspective. The positions and times measured by Bob are different, and from his perspective, Alice is accelerating away from him (green trajectory). The first number in the time superscript indicates the trigger time by Alice and the arrival time at Bob, and the second number indicates the explosion number.} \label{FIGURE001b}
\end{figure}

For simplicity, we assume again that Alice and Bob begin at rest at the origins of their respective coordinate systems and that the coordinate axes of both observers overlap at an initial time $t_{\rm A} = t_{\rm B} = 0$. In addition, As illustrated in Fig.~\ref{FIGURE001a}, we first study the case where Bob moves in the direction of the positive $x$ axis away from Alice and Alice sends light pulses with $s=1$ to Bob. In Alice's reference frame, the separation between two successive light lines in terms of her natural coordinates $\chi_{\rm A}$ is given by $\Delta \chi_{\rm A} = -c\Delta t_{\rm A}$ which Bob receives at intervals of $\Delta \chi_{\rm B}(\chi_{\rm A})$. The intervals recorded by Bob between each received narrow light pulse are in general, not the same because his speed varies constantly with respect to Alice's reference frame. As illustrated in Fig.~\ref{FIGURE001b}, he experiences space and time in a constantly changing way. Only the speed of light remains constant. In the following, we denote the natural coordinates of the light line created by the $N$-th explosion by $\chi_{\rm A}^{(N)}$ and $\chi_{\rm B}^{(N)}$. Using this notation, we know that
\begin{eqnarray} \label{sk1}
    \chi_{\rm A}^{(N)} = N \, \Delta \chi_{\rm A} ~~ \rm{and} ~~
    \chi_{\rm B}^{(N)} = \sum_{n=1}^{N} \Delta \chi_{\rm B}^{(n)} 
\end{eqnarray} 
with $\Delta \chi_{\rm B}^{(n)} = \chi_{\rm B}^{(n)} - \chi_{\rm B}^{(n-1)}$. In Figs.~\ref{FIGURE001a} and \ref{FIGURE001b} respectively, $\Delta \chi_{\rm A}$ and $\Delta \chi_{\rm B}^{(n)}$ coincide with the lengths of the bottom edges of the light blue triangles. 

In the following, we denote the times when Alice triggers the $n^{\rm th}$ explosion by $t_{{\rm A}}^{(1,n)}$ and $t_{{\rm B}}^{(1,n)}$  in Alice's and Bob's frames respectively, while $t_{{\rm A}}^{(2,n)}$ and $t_{{\rm B}}^{(2,n)}$ mark the arrival times of the narrow light pulses at Bob's position.
Since, in each segment, Bob moves away from Alice at a constant speed, a closer look at Figs.~\ref{FIGURE001a} and \ref{FIGURE001b} suggests that, in analogy to Eq.~(\ref{sk0}),
\begin{eqnarray} \label{sk5}
&& \Delta \chi_{\rm A} = - c \, \Delta t_{{\rm A}}^{(1,n)} = - ( c - v_{\rm B} (\chi_{\rm A})) \Delta t_{{\rm A}}^{(2,n)}  \, , \nonumber\\
&& \Delta \chi_{\rm B}^{(n)} = - (c +  v_{\rm B} (\chi_{\rm A})) \Delta t_{{\rm B}}^{(1,n)} = - c \, \Delta t_{{\rm B}}^{(2,n)} ~~~ 
\end{eqnarray}
with $\Delta t_{i}^{(m,n)} \equiv t_{i}^{(m,n)} -  t_{i}^{(m,n-1)}$ for $m=1,2$. In addition, the principle of relativity now implies that  
\begin{eqnarray} \label{final}
{\Delta t_{{\rm B}}^{(1,n)} \over \Delta t_{{\rm A}}^{(1,n)}} = {\Delta t_{{\rm A}}^{(2,n)} \over \Delta t_{{\rm B}}^{(2,n)}} = \gamma \big(\chi_{\rm A}^{(n)} \big) ~~~
\end{eqnarray} 
in analogy to Eq.~(\ref{dilation}). The factor $\gamma \big(\chi_{\rm A}^{(n)} \big)$ in this equation characterises the time dilation between subsequent
light lines which depends only on the current relative speed $v_{\rm B} (\chi_{\rm A}))$ of Bob with respect to Alice. Proceeding as in the previous subsection when combining Eqs.~(\ref{sk5}) and (\ref{final}), we therefore now find that 
\begin{eqnarray} \label{oma}
\Delta \chi_{\rm B}^{(n)} &=& \gamma(\chi_{\rm A}^{(n)}) \left( 1+ \beta(\chi_{\rm A}^{(n)}) \right) \Delta \chi_{\rm A}  \, ,
\end{eqnarray}
with the constants $\beta(\chi^{(n)}_{\rm A})$ and $\gamma \big(\chi_{\rm A}^{(n)} \big) $ defined such that 
\begin{eqnarray} \label{sevi}
\beta(\chi^{(n)}_{\rm A}) &=& v_{\rm B}(\chi_{\rm A}^{(n)}) / c \, , \nonumber\\
\gamma \big(\chi_{\rm A}^{(n)} \big) &=& \left[ 1 - \beta \big(\chi_{\rm A}^{(n)} \big)^2 \right]^{-1/2} . 
\end{eqnarray}
Not surprisingly, the above equations show that the transformation between $\Delta \chi_{\rm A}$ and $\Delta \chi_{\rm B}^{(n)}$ for each individual line segment is nothing more than the relevant Doppler coefficient (cf.~Eq.~(\ref{S01})). 

If we moreover consider the analogous scenario where Bob sends narrow light pulses to Alice, as in the derivation of Eq.~(\ref{S01}), take the limit of small $\Delta \chi_{\rm A}$ and $\Delta \chi_{\rm B}$ respectively and substitute the above result into Eq.~(\ref{sk1}), we now find that
\begin{eqnarray}
    \label{n011}
    \chi_{\rm B} &=& \int_0^{\chi_{\rm A}} {\rm d}\chi_{\rm A}' \, \gamma(\chi_{\rm A}') \left( 1+ s \beta(\chi_{\rm A}') \right)
\end{eqnarray}
in the case of a stationary and a general non-inertial observer, Alice and Bob, who are placed at the origins of their respective coordinate systems and overlap at an initial time $t_{\rm A}=t_{\rm B}=0$. 
Eq.~(\ref{n011}) is the main result of this subsection and shows how the natural coordinates $\chi_{\rm A}$ and $\chi_{\rm B}$ which Alice and Bob use to describe the same light line relate to each other. Compared to 
non-inertial reference frames, the coefficients $\beta $ and $\gamma$ have now been replaced by position-dependent functions $\beta(\chi_{\rm A}')$ and $\gamma(\chi_{\rm A}')$ where
\begin{eqnarray} \label{oma3}
\beta(\chi_{\rm A}') = v_{\rm B}(\chi_{\rm A}')/c \, , ~~ 
\gamma(\chi_{\rm A}') = \left[1 - \beta(\chi_{\rm A}')^2 \right]^{-1/2}
\end{eqnarray}
which can be calculated analytically for any given trajectory of Bob with respect to Alice's spacetime diagram before reaching his current position, i.e.~when Bob's velocity $v_{\rm B}(\chi_{\rm A}')$ relative to Alice whenever he passes a light-like trajectory with the natural coordinate $\chi_{\rm A}'$ is known.

\section{Results} 

Now we can discuss the implications of the transformation between $\chi_{\rm A}$ and $\chi_{\rm B}$ in Eq.~(\ref{n011}) for the Unruh effect. As pointed out earlier, the blip annihilation operators $a_{s\lambda}(\chi_{\rm A})$ and $b_{s\lambda}(\chi_{\rm B})$ of Alice and Bob are  associated with light-like trajectories or rays in the spacetime diagram. Since we also know that $s$ and $\lambda$ are invariant under reference frame transformations, we can fix $s$ and $\lambda$ in the following considerations without losing generality. As each light-like trajectory is common to all observers, we can use the conservation of the blip worldlines to establish a relationship between $a_{s\lambda}(\chi_{\rm A})$ and $b_{s\lambda}(\chi_{\rm B})$.  To do so, we consider a single fixed region of spacetime with a fixed number $N$ of light-like trajectories passing through. Although this ``box" contains the same amount of light in both reference frames, its width $\Delta\chi_{\rm A}$ and $\Delta\chi_{\rm B}$ in Alice's and Bob's reference frames is not the same for both observers.

More concretely, the number of blips in the ``box" is the same for both observers when
\begin{eqnarray}
	\label{Nlight}
N &=& \int_{\Delta \chi_{\rm B}} {\rm d}\chi_{\rm B}\, b^\dagger_{s\lambda}(\chi_{\rm B})b_{s\lambda}(\chi_{\rm B}) \nonumber \\
&=& \int_{\Delta \chi_{\rm A}} {\rm d}\chi_{\rm A}\, a^\dagger_{s{\sf \lambda}}(\chi_{\rm A})a_{s\lambda}(\chi_{\rm A})\, .
\end{eqnarray}
This equivalence is satisfied for a ``box" of any size only when a local Doppler factor is introduced and added to the respective blip operator. We can demonstrate this result by substituting ${\rm d}\chi_{\rm B}/{\rm d}\chi_{\rm A}$ derived from Eq.~(\ref{n011}) into Eq.~(\ref{Nlight}). Moreover, taking into account that
\begin{eqnarray}
\label{Nlight2}
N &=& \int_{\Delta \chi_{\rm B}} \left(\frac{{\rm d}\chi_{\rm B}}{{\rm d}\chi_{\rm A}}\right){\rm d}\chi_{\rm A}\, b^\dagger_{s\lambda}(\chi_{\rm B})b_{s\lambda}(\chi_{\rm B})\, ,
\end{eqnarray}
we obtain the transformation
\begin{eqnarray}
\label{E45}
b_{s\lambda}(\chi_{\rm B}) &=& \sqrt{\gamma(\chi_{\rm A}) \big( 1- s \beta(\chi_{\rm A})\big)} \, a_{s\lambda}(\chi_{\rm A})
\end{eqnarray}
which is identical to the transformation in Eq.~(45) in Ref.~\cite{Hodgson:2024mfs}. Like the $a_{s\lambda}(\chi_{\rm A})$ annihilation operators, one can now check that the $b_{s\lambda}(\chi_{\rm B})$ obey the bosonic commutation relation 
\begin{eqnarray}
\label{comm11}
\big[b_{s\lambda}(\chi_{\rm B}),b^{\dagger}_{s^{\prime}\lambda^{\prime}}(\chi_{\rm B}')\big]
&=& \delta_{ss\prime}\,\delta_{\lambda\lambda^{\prime}}\,\delta(\chi_{\rm B}-\chi_{\rm B}') \, ,
\end{eqnarray}
which is analogous to Eq.~(\ref{comm1}), as expected. Hence we could also have obtained Eq.~(\ref{E45}) in a more direct way by demanding that blips remain local in all reference frames and imposing bosonic commutator relations on the $b_{s \lambda}(\chi_{\rm B})$ annihilation operators.

The transformation in Eq.~(\ref{E45}) is the main result of this paper 
and analogous to the transformation obtained in the Doppler case where Bob moves at a constant velocity through Alice's reference frame \cite{Hodgson:2024mfs}. As mentioned already before, a moving and an accelerating observer experience space and time differently and therefore use different natural coordinates when describing local field excitations, but otherwise, nothing changes. Notice also that Eq.~(\ref{E45}) can now be used to calculate the quantum state $|\psi_{\rm B} \rangle$ of the EM field in the moving frame for any given quantum state $|\psi_{\rm A} \rangle$ in the resting frame simply by substituting the $b_{s\lambda}(\chi_{\rm B})$ operators for the $a_{s\lambda}(\chi_{\rm A})$ operators. With respect to the Unruh effect \cite{Unruh,Unruh2,Unruh3}, the key result to note is our prediction that Alice and Bob share a common vacuum, which contrasts with earlier investigations \cite{Berra-Montiel:2016xnn,Volovik:2022cqk,Kialka:2017ubk}. The only difference between the inertial and the general case is that the light-like worldline densities, i.e.~the number of worldlines $N$ in a given region divided by the $V$ volume that they occupy, are only homogeneous in space when Bob moves with constant velocity \cite{Hodgson:2024mfs}. 
Whether the standard description or our closely related local photon approach provides a more accurate description of the quantised EM field eventually needs to be decided by experiments.

\section{Discussion}

Eq.~(\ref{E45}) shows that transforming  local blip annihilation operators $a_{s \lambda}(\chi_{\rm A})$ between non-inertial reference frames only adds an overall factor, which can be explained by the fact that observers in different reference frames experience space and time differently.
What is more, the number of excitations remains constant under reference frame transformations. In other words, an observer in an inertial and in a non-inertial reference frame both experience frequency shifts of the light that they are observing, as in the case of the relativistic Doppler effect \cite{Hodgson:2024mfs}. To better illustrate the findings in this paper, in this section, we now consider the familiar example where Bob travels with a constant acceleration with respect to Alice's reference frame. In the literature, this specific scenario and not the general scenario which we considered in the previous sections is usually associated with the Unruh effect \cite{Unruh,Unruh2,Unruh3}.

\subsection{Uniform acceleration} 

In the following, we therefore assume that, according to Alice who is at rest at $x_{\rm A} = 0$, Bob experiences a constant accelerating force and follows a trajectory defined by the velocity 
\begin{eqnarray}
\label{velocity1}
v_{\rm B}(t_{\rm A}) &=& \frac{at_{\rm A}}{\sqrt{1+(at_{\rm A}/c)^2}} 
\end{eqnarray}
with $v_{\rm B}(0) = 0$. This velocity increases asymptotically towards but never exceeds the speed of light $c$. Assuming again that when Alice's clock measures $t_{\rm A} = 0$ Bob is stationary at $x_{\rm A}(0) = 0$, his position at time $t_{\rm A}$ equals 
\begin{eqnarray}
\label{int_1}
x_{\rm A}(t_{\rm A}) &=& \int_{0}^{t_{\rm A}}{\rm d}t'_{\rm A}\;v_{\rm B}(t'_{\rm A}) \, .
\end{eqnarray}
Using Eq.~(\ref{velocity1}), we therefore find that Bob's position in Alice's frame is given by
\begin{eqnarray}
\label{int_11}
x_{\rm A}(t_{\rm A}) &=& \frac{c^2}{a}\left[\sqrt{1+(at_{\rm A}/c)^2} - 1\right]\, .
\end{eqnarray}
In the limit of relatively small accelerations and for sufficiently short times, i.e.~for $a t_{\rm A}/c \ll 1$, the above expressions simplify to $v_{\rm B}(t_{\rm A}) = at_{\rm A}$ and $x_{\rm A}(t_{\rm A}) = a t_{\rm A}^2/2$, which is the non-relativistic velocity and position of a uniformly accelerating observer. 

\begin{figure}[t]
\centering
\includegraphics[width = 0.45\textwidth]{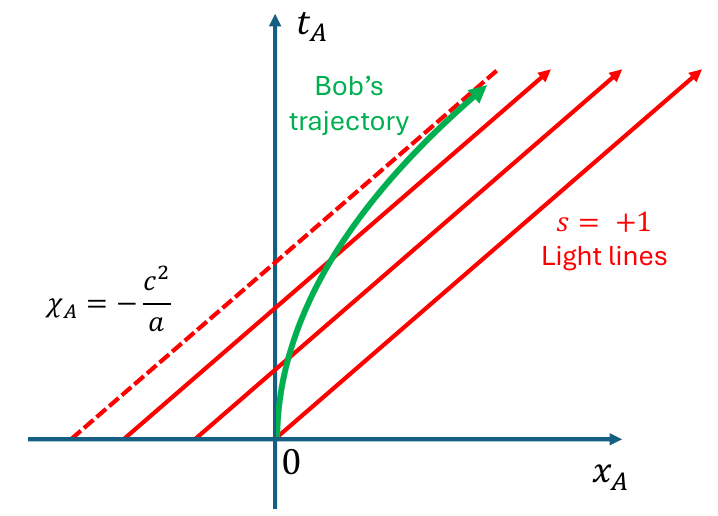}
\caption{The figure shows Bob's trajectory in Alice's spacetime diagram with perpendicular axes representing Alice's position $x_{\rm A}$ and time $t_{\rm A}$. The green line shows Bob's hyperbolic trajectory in Alice's reference frame. Right-propagating light pulses, shown in red, only intersect with Bob's trajectory if they have a $\chi_{\rm A}$ coordinate in the range $\chi_{\rm A} \in (-c^2/a, 0]$. Left-propagating light pulses will only intersect with Bob's trajectory if $\chi_{\rm A} \geq 0$. The dashed light pulse at $\chi_{\rm A} = -c^2/a$ meets Bob at a time $t_{\rm A} \to \infty$ and therefore provides a natural horizon for the $\chi_{\rm A}$ and $\chi_{\rm B}$ coordinate systems.}
\label{Fig:Rindler}
\end{figure}

As we have seen in the previous section, to determine how the state of the quantised EM field changes under reference frame transformations, we need to know $v_{\rm B}$ not as a function of $t_{\rm A}$ but as a function of the natural coordinate $\chi_{\rm A}$. To calculate Bob's velocity when crossing the light line with the natural coordinate $\chi_{\rm A}$ in Alice's reference frame, we now have a closer look at Fig.~\ref{Fig:Rindler}, which shows that 
\begin{eqnarray} \label{oma2}
\chi_{\rm A}(t_{\rm A}) &=& x_{\rm A}(t_{\rm A}) - sct_{\rm A} 
\end{eqnarray}
with $x_{\rm A}(t_{\rm A})$ given in Eq.~(\ref{int_11}). Combining Eqs.~(\ref{int_11}) and (\ref{oma2}) and calculating $t_{\rm A}$ in terms of $\chi_{\rm A}$ for the points along Bob's trajectory yields
\begin{eqnarray}
\label{velocity2}
t_{\rm A}(\chi_{\rm A}) &=& - \frac{s \chi_{\rm A}}{c} \cdot \frac{1+a\chi_{\rm A}/2c^2}{1+a\chi_{\rm A}/c^2} 
= {sc \over 2a} \cdot {1- \eta(\chi_{\rm A})^2 \over \eta(\chi_{\rm A})} \nonumber \\
\end{eqnarray}
with $\eta(\chi_{\rm A}) = 1+ a \chi_{\rm A}/c^2$. Hence, Eq.~(\ref{velocity1}) can be used to show that Bob's velocity $v_{\rm B}(\chi_{\rm A})$ is given by
\begin{eqnarray}
\label{velocity2}
v_{\rm B}(\chi_{\rm A}) &=& sc \cdot \frac{1-\eta(\chi_{\rm A})^2}{1+\eta(\chi_{\rm A})^2} \, .
\end{eqnarray}
As illustrated in Fig.~\ref{Fig:Rindler}, for $v_{\rm B} > 0$ the relevant natural coordinates $\chi_{\rm A}$ lie within the range $\chi_{\rm A} \in (-c^2/a, 0]$ when $s=1$. Analogously, one can show that $\chi_{\rm A} \geq 0$ when $s=-1$. Light signals originating from sources located beyond this range are beyond Bob's event horizon and cannot be observed by Bob. Therefore, $v_{\rm B}(\chi_{\rm A})$ is not defined within these regions.

Now we can use Eq.~(\ref{n011}) to obtain the transformation between the natural coordinates $\chi_{\rm A}$ and $\chi_{\rm B}$ used by Alice and Bob to describe the same narrow light pulse within their respective reference frames. By substituting the velocity (\ref{velocity2}) into Eq.~(\ref{n011}), one finds that
\begin{eqnarray}
\label{Rindler0}
\chi_{\rm B}(\chi_{\rm A}) 
= \int_{0}^{\chi_{\rm A}} {\rm d} \chi_{\rm A}' \, \frac{1}{\eta(\chi_{\rm A}')}
= \frac{c^2}{a}\ln \left| \eta(\chi_{\rm A}) \right|\, .
\end{eqnarray}
Notice that these coordinates are only valid for light-like trajectories defined within the permitted regions $\chi_{\rm A} \in (-c^2/a, 0]$ for right-propagating light and $\chi_{\rm A} \geq 0$ for left-propagating light. Unlike in the case of uniform velocity \cite{Hodgson:2024mfs}, the relationship between $\chi_{\rm A}$ and $\chi_{\rm B}$ is strongly non-linear. Interestingly, the length transformation in Eq.~(\ref{Rindler0}) has many similarities with the inverse frequency transformation of a uniformly accelerating observer, which has been derived in Refs.~\cite{Lin,3,4} based on a {\em hypothesis on locality}.

Finally, combining Eqs.~(\ref{velocity2}) and (\ref{Rindler0}) with the main result of this paper, namely the transformation between the blip annihilation operators $a_{s\lambda}(\chi_{\rm A})$ and $b_{s\lambda}(\chi_{\rm B})$ in Eq.~(\ref{E45}), we find that 
\begin{eqnarray} \label{C39}
b_{s\lambda}(\chi_{\rm B}) &=& \left| \eta(\chi_{\rm A}) \right|^{1/2} \,a_{s\lambda}(\chi_{\rm A}) \nonumber \\
&=& {\rm e}^{a \chi_{\rm B}/2 c^2} \, a_{s\lambda}(\chi_{\rm A})
\end{eqnarray}
for the case of the uniformly accelerating observer, which we introduced in~Eq.~(\ref{velocity1}). For completeness, we now also have a closer look at the transformation of the annihilation operators $\widetilde a_{s\lambda}(k_{\rm A})$ and $\widetilde b_{s\lambda}(k_{\rm B})$ in momentum space where $k_{\rm A}$ and $k_{\rm B}$ are the wave numbers of the photons in Alice's and Bob's reference frames, respectively. Applying the Fourier transforms in Eq.~(\ref{mom1}) to Eq.~(\ref{C39}), we find that
\begin{eqnarray} \label{C40}
\widetilde{b}_{s\lambda}(k_{\rm B})
&=& \frac{1}{2\pi} 
\int_{-\infty}^{\infty} {\rm d}k_{\rm A} 
\int_{-\infty}^{\infty} {\rm d}\chi_{\rm B}(\chi_{\rm A})
\, {\rm e}^{a \chi_{\rm B}(\chi_{\rm A})/2c^2} \nonumber \\
&& \times  
{\rm e}^{{\rm i}s (k_{\rm A} \chi_{\rm A} - k_{\rm B} \chi_{\rm B}(\chi_{\rm A}))} \; 
\widetilde{a}_{s\lambda}(k_{\rm A}) \, .
\end{eqnarray}
This relation shows that monochromatic waves in Bob's reference frame correspond to complex photonic wave packets in Alice's reference frame. This is not surprising since the variable transformation in Eq.~(\ref{Rindler0}) is not homogeneous and local field excitations get deformed differently everywhere (cf.~Eq.~(\ref{C39})). However, the vacuum state is the same everywhere in all cases.

\subsection{Bogoliubov transformations} 

In contrast to this, transformations between inertial and uniformly accelerating reference frames usually map the vacuum state onto a thermal state \cite{Cri}. A common way of arriving at this effect, i.e.~the Unruh effect \cite{Unruh,Unruh2,Unruh3}, is to assume that the expectation values of the quantised EM field change in agreement with special relativity and with classical quantum electrodynamics \cite{Maybee}. To achieve this, the annihilation operators of monochromatic photons need to undergo a Bogoliubov transformation, which changes annihilation operators into a mix of annihilation and creation operators which leads to the prediction of particles in the accelerating frame, even when the quantised EM field appears to be in its vacuum state in the inertial frame. In the standard formulation of quantum field theory, the Unruh effect is simply an unavoidable mathematical necessity. 

To better understand how our local photon approach avoids the mixing of creation and annihilation operators in the reference frame of the accelerating observer, let us return to the discussion in Section \ref{oma4}. In the standard description of the quantised EM field, all particles evolve according to a strictly positive Hamiltonian. In momentum space, the expectation values of annihilation and creation operators therefore accumulate phase factors of the form ${\rm exp}(-{\rm i}\omega t)$ and ${\rm exp}({\rm i}\omega t)$ with $\omega \ge 0$, respectively, as time progresses. In contrast to this, our local photon approach supports a larger Hilbert space \cite{Jake,Hodgson2022,Gabe} and the dynamical Hamiltonian has both positive and negative eigenvalues (cf.~Eq.~(\ref{Hdyn2})). As a result, the expectation values of the photon annihilation operators $a_{s \lambda}(k)$ now accumulate phase factors of the form ${\rm exp}(-{\rm i}\omega t)$ with $\omega \in (-\infty,\infty)$. It is therefore possible to transform the expectation values of electric and magnetic field observables in agreement with special relativity and with classical quantum electrodynamics {\em without} invoking Bogoliubov transformations, but simple linear transformation, like the one shown in Eq.~(\ref{C40}).

\subsection{Detector dependence}

Formally, Bogoliubov transformations are still allowed in our model. However, as this paper tries to emphasize, the annihilation operators which we obtain via such a transformation might not represent actual photons. Suppose the detector resembles a localised two-level system, i.e.~a so-called Unruh-DeWitt detector, with ground state $|0 \rangle$ and excited state $|1 \rangle$ \cite{Unruh2,UDD,UDD2,UDD3}. Then there are many ways of representing the Hamiltonian $H$ of the total system \cite{Adam,Adam2}. This applies, since it is always possible to replace the Hamiltonian $H$ of the total system by a Hamiltonian 
\begin{eqnarray}
H' &=& U \, H \, U^\dagger 
\end{eqnarray}
with $U^\dagger U = U U^\dagger = 1$ without changing the spectrum. However, each of the above representations corresponds to a different decomposition of the composite quantum system of detector and field into subsystems and to different forms of the interaction Hamiltonian. For example, $U$ transforms the field annihilation operators $a_{s \lambda}(x)$ into
\begin{eqnarray}
a_{s \lambda}(x)' &=& U \, a_{s \lambda}(x) \, U^\dagger 
\end{eqnarray}
which represents a different physical object.
Despite the uniqueness of the predictions for the dynamics of the total system, the uniqueness of the predictions for subsystems cannot be guaranteed. 

However, the above described ambiguity, i.e.~the gauge-dependence, of the definition of subsystems can be removed using arguments from thermodynamics. As shown in Ref.~\cite{Adam}, consistency with the second law of thermodynamics, i.e.~the impossibility of heat flow from a colder to a hotter system, requires that the interaction Hamiltonian between the two-level system and the quantised EM field commutes with the free energy. Moreover, demanding locality of the interactions between the detector and
the EM field, we find that their interaction Hamiltonian $H_{\rm int}$ must take the form
\begin{eqnarray} \label{C41}
H_{\rm int} &=& \sum_{s =\pm 1} \hbar g \, |0 \rangle \langle 1| \, a_{s \lambda}^\dagger(x) + {\rm H.c.} 
\end{eqnarray}
where $g$ denotes a coupling constant and where $x$ specifies the position of the detector \cite{Tom}. In other words, the operators $|0 \rangle \langle 1|$ and $ a_{s \lambda}^\dagger(x)$ only refer to detector and field, respectively, if their interaction is of the above form. Most importantly, the above interaction only excites the detector in the presence of actual photons. Considering a two-level atom in free space in the absence of external laser driving with realistic experimental parameters and using 
and an unphysical interaction Hamiltonian can result in the predictions of stationary photon emission rates of $4 \times 10^6$ photons per second for realistic experimental parameters \cite{Kurcz1,Kurcz2}. This prediction does not even require the acceleration of the source and nevertheless contradicts actually observed photon numbers by many orders of magnitude.

\section{Conclusions}

Recently, it was shown that replacing the usual canonical quantisation of the EM field \cite{Birrell} by a physically motivated quantisation of light in position space \cite{Jake,Hodgson2022,Gabe} leads to a doubling of the Hilbert space of the quantised EM field. The expanded state space supports local photon annihilation and creation operators $a_{s\lambda}(\chi)$ and $a^\dagger_{s\lambda}(\chi)$ with natural coordinates $\chi = x - sct$. The blip annihilation operators which describe the same light-like trajectory in different non-inertial reference frame relate to each other in position space via a local transformation (cf.~Eq.~(\ref{A2})) which is formally the same as the transformation between observers with constant relative speed \cite{Hodgson:2024mfs}. As in the Doppler effect, our theory predicts changes in the local density of light-like worldlines in the reference frame of moving observers (cf.~Eq.~(\ref{E45})). Moreover, the vacuum remains the same in all inertial and non-inertial reference frames. This prediction is in contrast to the standard photon pair creation result of an observer with constant acceleration, which became known as the Unruh effect \cite{Unruh,Unruh2,Unruh3,Cri}. Our approach is likely to stimulate further research into quantum electrodynamics in non-inertial reference frames and emphasizes the need for more conclusive experiments.

\vspace*{0.5cm}
\noindent
{\em \bf Acknowledgement.} A.B. and D.H. would like to thank Robert Purdy for stimulating discussions. D.H. acknowledges financial support from the UK Engineering and Physical Sciences Research Council EPSRC [grant number EP/W524372/1].


\end{document}